# Prediction of Shared Bicycle Demand with Wavelet Thresholding

*J. Christopher Westland: westland@uic.edu, University of Illinois – Chicago, USA*

*Jian Mou: jian.mou@xidian.edu.cn, XiDian University, China*

*Dafei Yin: yindafei@mobike.com, Mobike, China*

*February 7, 2018*

# Abstract

Consumers are creatures of habit, often periodic, tied to work, shopping and other schedules. We analyzed one month of data from the world's largest bike-sharing company to elicit demand behavioral cycles, initially using models from animal tracking that showed large customers fit an Ornstein-Uhlenbeck model with demand peaks at periodicities of 7, 12, 24 hour and 7-days. Lorenz curves of bicycle demand showed that the majority of customer usage was infrequent, and demand cycles from time-series models would strongly overfit the data yielding unreliable models. Analysis of thresholded wavelets for the space-time tensor of bike-sharing contracts was able to compress the data into a 56-coefficient model with little loss of information, suggesting that bike-sharing demand behavior is exceptionally strong and regular. Improvements to predicted demand could be made by adjusting for 'noise' filtered by our model from air quality and weather information and demand from infrequent riders.

*(147 words)*

Consumers are creatures of habit, and this trait has long been an important strategic tool in branding, logistics and demand forecasting. These habits reinforce cycles of behavior that are often periodic, tied to work and shopping schedules, seasons, and hours of the day. Forecasting and analysis of consumer demand are commonly represented as stable patterns of behavior over a time-series.

In their comprehensive paper (Aghabozorgi, Shirkhorshidi et al. 2015) argued that time-series representations can classified in four ways: (1) nondata adaptive; (2) data adaptive; (3) model-based; and (4) data dictated. Time-series for stable patterns of demand are nondata adaptive representations in *Aghabozorgi, et al's* taxonomy and their transformation parameters remain the same for all time-series. The best-known methods for nondata adaptive representations are piecewise aggregate approximation, discrete wavelet transform, discrete Fourier transform, discrete cosine transform and perceptually important points. We proposed and evaluated a specific application of thresholding for discrete wavelet transforms in this paper for the forecasting of future demand in bicycle sharing markets.

We address the following specific research objective: identify human behavioral habits reflected in demand cycles which can be used for the improved geographical positioning of Mobike bikes in a predefined service region at a specific point in future time. Demand and supply management is of central importance to bike sharing companies, because demand can only be satisfied when the bicycle assets are available at the same specific place and time as the customer. Logistics at Mobike is primarily focused leveling geographical excesses of supply and demand for bikes using (1) trucks to move bikes around geographically; (2) purchase of more bikes from vendors to fill deficiencies; or (3) rewarding riders to pedal bikes to locations where demand is high (what Mobike calls their "red packet" program after the monetary envelope given during Asian holidays).

Precise economic figures on bike sharing revenues do not exist, but have been estimated to approach $10 billion, growing at 20% annually at the end of 2017 (El-Assi, Mahmoud et al. 2017). It is believed that most of the world's bike sharing companies appear to be operating at a loss, as competition is intense. Increasing revenues combined with ongoing losses suggest that operational improvements are dearly needed at bike sharing companies. Most improvements to

logistics and supply-demand rebalancing depend on accurate forecasts of future demand, and this will be the focus of the current research.

## Methods

### Mobike's Bicycle Sharing

Our researchers were provided a large dataset of usage statistics from the World's largest shared bicycle operator, Beijing Mobike Technology Co., Ltd. (北京摩拜科技有限公司) a fully station-less bicycle-sharing system headquartered in Beijing, China. The firm is currently valued at US$3 billion (Steinberg 2017). The majority (76.22%) of data points lie around $[longitude, latitude] \cong [116.4, 39.9]$ which is close to, but a bit East of Mobike Headquarters at $[longitude, latitude] \cong [116.2, 39.9]$. We have consequently focused on Beijing data, classified as being areas roughly within the 5$^{th}$ ring road of Beijing, and between latitudes $[39.8, 40.0]$ and longitudes $[116.0, 116.8]$. The correlation between frequent usage and average distance, or average usage time is low, and it was generally difficult to discern any particular order from the distance or usage time from the data; these values were generally small, indicating that Mobikes are used for short trips and are a convenient alternative to walking.

|  | *Worldwide* | *Beijing* |
|---|---|---|
| *Total Number of Contracts* | 73,380,988 | 55,928,760 |
| *Unique Bikes* | 1,169,113 | 657,346 |
| *Unique Customers* | 4,982,788 | 3,981,765 |

|  | *Worldwide* | | | | | | | *Beijing* | | | | |
|---|---|---|---|---|---|---|---|---|---|---|---|---|
|  | *Start longitude* | *Start latitude* | *End longitude* | *End latitude* | *contract dates* | *contract length* | *contract distance (degrees)* | *Start longitude* | *Start latitude* | *contract dates* | *contract length* | *contract distance (degrees)* |
| *Min.* | -118.5 | -0.006 | -123.3 | -33.95 | 7/15/17 | 0 | 0 | 116 | 39.8 | 7/15/17 | 0S | 0 |
| *1st* | 116.3 | 39.871 | 116.3 | 39.87 | 7/22/17 | 4M 44S | 0.00453 | 116.3 | 39.88 | 7/22/17 | 4M 39S | 0.00453 |
| *Median* | 116.4 | 39.918 | 116.4 | 39.92 | 7/29/17 | 7M 39S | 0.00788 | 116.4 | 39.91 | 7/29/17 | 7M 26S | 0.00774 |
| *Mean* | 116.4 | 39.858 | 116.3 | 39.82 | 7/29/17 | 11M 53S | 0.1278 | 116.4 | 39.91 | 7/29/17 | 11M 27.2 | 0.12431 |
| *3rd* | 116.5 | 39.97 | 116.5 | 39.97 | 8/4/17 | 13M 1S | 0.0138 | 116.5 | 39.95 | 8/4/17 | 12M 36S | 0.01338 |
| *Max.* | 135.5 | 53.48 | 151.3 | 53.48 | 8/13/17 | 7d 17H 10M 28S | 239.9688 | 116.8 | 40 | 8/13/17 | 7d 17H 10M 28S | 239.8812 |

*Figure 1: Summary of Mobike's Data*

*Prior Socio-economic-demographic Bike-Sharing Research*

There is an established and broadly consistent body of evidence about riders' socio-economic and demographic profile. In general, bike sharing services attract a population that is: male, white, employed; and compared to the average population in which sharing services are implemented, younger, more affluent, more educated and more likely to be already engaged in cycling independently of bike sharing (Shaheen, Guzman et al. 2010, Shaheen and Guzman 2011, Shaheen, Zhang et al. 2011, Shaheen, Guzman et al. 2012, Fishman, Washington et al. 2014).

In Beijing, Shanghai and Hangzhou, bike sharing users were found to have a higher level of car ownership than non-users (Shaheen, Zhang et al. 2011)but on average gender, class and ethnic differences bike sharing reflect general demographics of bike usage. Accessibility and coverage are important in the success of bike sharing programs (Shaheen, Guzman et al. 2012)and trip rates were higher in poorer areas, suggesting that bike sharing services may provide the primary means of transportation.

*Policy Models for Bike Sharing Management*

Substantial research has been dedicated to ride sharing analysis for cars, and though there is recent research in bike sharing, it is not nearly as extensive. There are notable differences. Both cars (with drivers) and bikes (with riders) are sentient, and thus can be incentivized (e.g., through surge pricing) to come together in a matching market that can be modeled with a bipartite network. But whereas cars are typically parked at restricted locations, Mobike's bikes can be "parked" anywhere that the last rider has left them. Self-driving cars could potentially return o where they are needed, but at least in the near future, bikes without riders will be non-sentient, and thus cannot be incentivized, nor do they move without intervention (e.g., from a rider, or from a Mobike truck which rebalances bike inventory on the 2D grid)

Research in bike sharing is quite recent, most appearing in the last decade. Studies have either been descriptive (DeMaio 2009, Shaheen, Guzman et al. 2010, Shaheen and Guzman 2011, Larsen 2013, Zhang, Zhang et al. 2015) or study bike sharing optimization problems (Vogel and Mattfeld 2010, Contardo, Morency et al. 2012, Fricker, Gast et al. 2012, García-Palomares, Gutiérrez et al. 2012, Raviv and Kolka 2013, Raviv, Tzur et al. 2013, Schuijbroek, Hampshire et al. 2013, Dell'Amico, Hadjicostantinou et al. 2014, Fricker and Gast 2016). Though optimization approaches depend implicitly on accurate dynamic rider demand elicitation, this is exogenous to their models, and sidestep what is perhaps the most difficult problem in bike inventory rebalancing – predicting rider demand and bicycle supply at a particular time and place. Failure to statistically analyze and predict demand and supply makes such models impractical with potentially misleading policy choices.

"Optimization" and "rebalancing" are intended to provide policy guides for the dynamic shifting of bicycle inventory (primarily by truck transport, but possibly through contract price incentives to riders) to position bicycles in space and time in a way that is consistent with rider demand. Their performance is predicated on the accurate prediction of rider demand at a particular point in geographical space (2D) and time. Data analysis of prior contracts and other information will provide the information to facilitate this accurate prediction. (Barnes and Krizek 2005) discussed the requirements of such a predictive model, but concluded that such a statistical model would be difficult to construct. Geospatial models in ecology is perhaps most advanced in providing demand models that can predict temporal-geospatial positions (Rossi, Mulla et al. 1992, Cravey, Washburn et al. 2001, Legendre, Dale et al. 2002, Griffith and Peres-Neto 2006). These models are intended for the analysis of consumer demand, but provide models that can inform our particular approach to the problem.

## *Wavelets Analysis Applied to Predicting Bike Sharing Demand*

This paper analyses a space-time representation of Mobike demand using on wavelet thresholding. This approach assumes that bicycle sharing demand data is contaminated with additive noise, injected by infrequent or irregular users of the sharing service, and that we analyze this data using a discrete wavelet transform (DWT). DWT coefficients are "thresholded" and noise coefficients discarded before we invoke the inverse wavelet transform to estimate the data and predict future demand. This is a common approach to compression and de-noising in image processing, which we finesse by replacing one of the spatial dimensions in 3D DWT with our time dimension.

A wavelet is a wave-like oscillation with an amplitude that begins at zero, increases, and then decreases back to zero. Wavelets can be combined, using "reverse, shift, multiply and integrate" convolutions, with portions of a known signal to extract information from the unknown signal. One can relate a wavelet decomposition to extracting notes from a musical tune. For example, a wavelet could be created to have a frequency of Middle C and a short duration of roughly a 32nd note. If this wavelet were to be convolved with a signal created from the recording of a tune, then the resulting signal would be useful for determining when the Middle C note was being played in the song. Mathematically, the wavelet will correlate with the signal if the unknown signal contains information of similar frequency. This concept of correlation is at the core of the demand signal extraction used in this research.

The basic idea of thresholding in the current bike demand 'signal plus noise' model assumes the wavelet transform of our bike location-time representation is very sparse but the wavelet transform of noise is not. The signal gets concentrated in few wavelet coefficients and the noise remains spread out, thus you can separate the signal from noise by keeping large wavelet coefficients and deleting the small. The smaller noise coefficients are presumed not to be important for future prediction (i.e., are transient, and due to irregular or infrequent users) and their inclusion in the model would overfit data and lead to erroneous predictions.

Wavelet thresholding acts as a filter to de-noise the space-time representation of Mobike demand. Hard and soft thresholding functions are defined in (Donoho and Johnstone 1994) by $\hat{d} = \eta_H(d^*, \lambda) = d^* I[|d^*| > \lambda]$ and $\hat{d} = \eta_S(d^*, \lambda) = \text{sgn}(d^*)(|d^*| - \lambda) I[|d^*| > \lambda]$ respectively, where $I$ is the indicator function, $d^*$ is the empirical coefficient to be thresholded, and $\lambda$ is the threshold. The most commonly used fit measure is integrated squared error $\widehat{M} = n^{-1} \sum_{i=1}^{n}[\hat{g}(x_i) - g(x_i)]^2$ where $\hat{g}(x_i)$ is the estimate and $\hat{g}(x_i) - g(x_i)$ the fit error. We have focused our performance metrics around these statistics, but use the more standard term "mean squared error" (MSE) rather than ISE in the subsequent analysis. Some models of fit refer to an "oracle" that is able to make "ideal" predictions, which we can then use to benchmark empirical fit. In practice we do not have access to a thresholding oracle, but (Donoho and Johnstone 1994) showed that if we perform wavelet shrinkage via soft thresholding with a threshold of $\sigma\sqrt{2 \cdot log(n)}$, the fit error comes within a log factor of the ideal fit, i.e., $\widehat{M} \leq (2 \cdot log(n) + 1)(\sigma^2 + M_{ideal})$. This is comparable or exceeds the fit of methods such as piecewise polynomial fits and variable-knot splines for estimation, but in general we do not have access to comparably precise non-wavelet methods for this analysis.

The current analysis uses the Daubechies least-asymmetric wavelets in the SureShrink algorithm developed in (Donoho and Johnstone 1995) using the (Stein 1981) unbiased risk estimation (SURE) technique. (Daubechies 1988) detailed the construction of orthogonal wavelets that are compactly supported but are smoother than traditional Haar wavelets via a solution of the dilation equation that resulted in a family of orthonormal indexed by the number of vanishing moments. This implements all of the previous concepts and is currently considered "best-practice" for wavelet thresholding. The primary resolution of the model is unique to a particular problem and comparable to choosing bandwidth in kernel regression on a logarithmic scale (Hall and Patil 1995, Hall and Nason 1997).

## Results

*Exploratory Analysis of Mobike Data*

Figure 2 provides a histogram of the rank-frequency of bike rentals during the research period. The histogram shows a sharp drop off between frequent and non-frequent renters, and this can be analyzed further using the Gini index and Lorenz curves. Gini coefficients are measures of the inequality among values of a frequency distribution and is the area between a 45-degree line and the Lorenz curve. The relationship of rider rank-frequency of sharing to inequality is depicted in figure 2 as well. A Gini coefficient of 1 expresses maximal inequality among values. It can be viewed as a measure of redundancy, lack of diversity, isolation, segregation, inequality, non-randomness, and compressibility. Note that the most frequent renters display much more consistent behavior than the entire population. Results for the entire population of 3,981,765 Beijing customers are likely biased by a large sub-population of casual renters with diverse behaviors. Figure 2 relies on several related metrics to show that Mobike customers are highly inhomogeneous. This in turn suggests the use of managerial strategies that trace activities and incentivizing high-frequency users. There is anecdotal evidence to support such strategies across the bike-sharing industry. Figures 2 shows that a strong "80-20 Rule" for ridership – a few riders are by far the heaviest users of Mobike contracts.

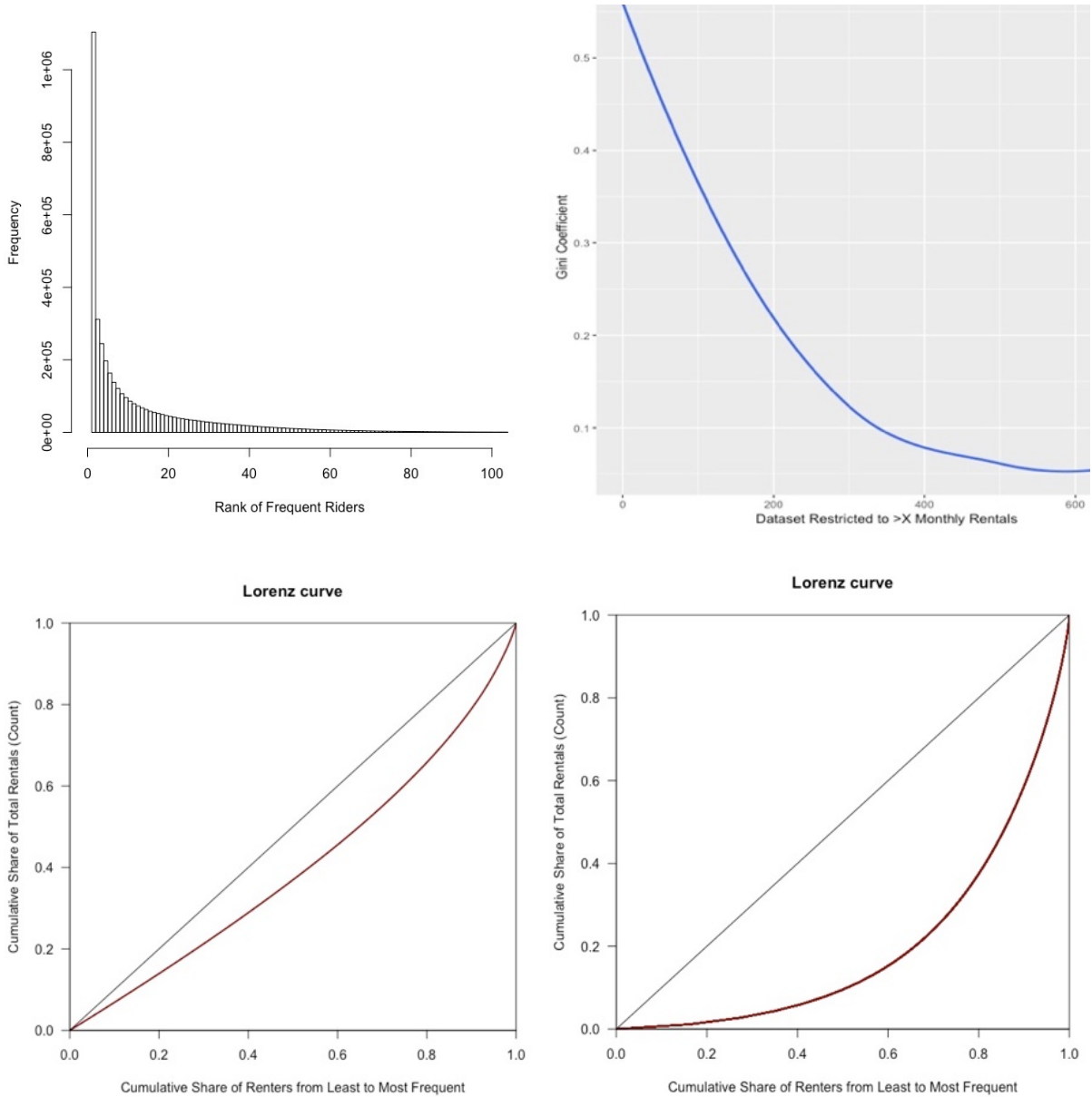

*Figure 2: Diversity (Inequality) of Rider Behavior; Histogram (**upper left**) of Rank-Frequency of Customer Contracts, fit to* $P(X \leq x) = \begin{cases} 1 - (24/x)^3 & \text{for } x \geq scale \\ 0 & elsewhere \end{cases}$ *; Gini Coefficient (**upper right**) Change in with Restriction to More Frequent Renters; (**lower left**) Lorenz Curve for All Renters; Gini 0.5991322; (**lower right**) Lorenz Curve for Renters Averaging More than 5 Rentals / Day; Gini 0.1977445*

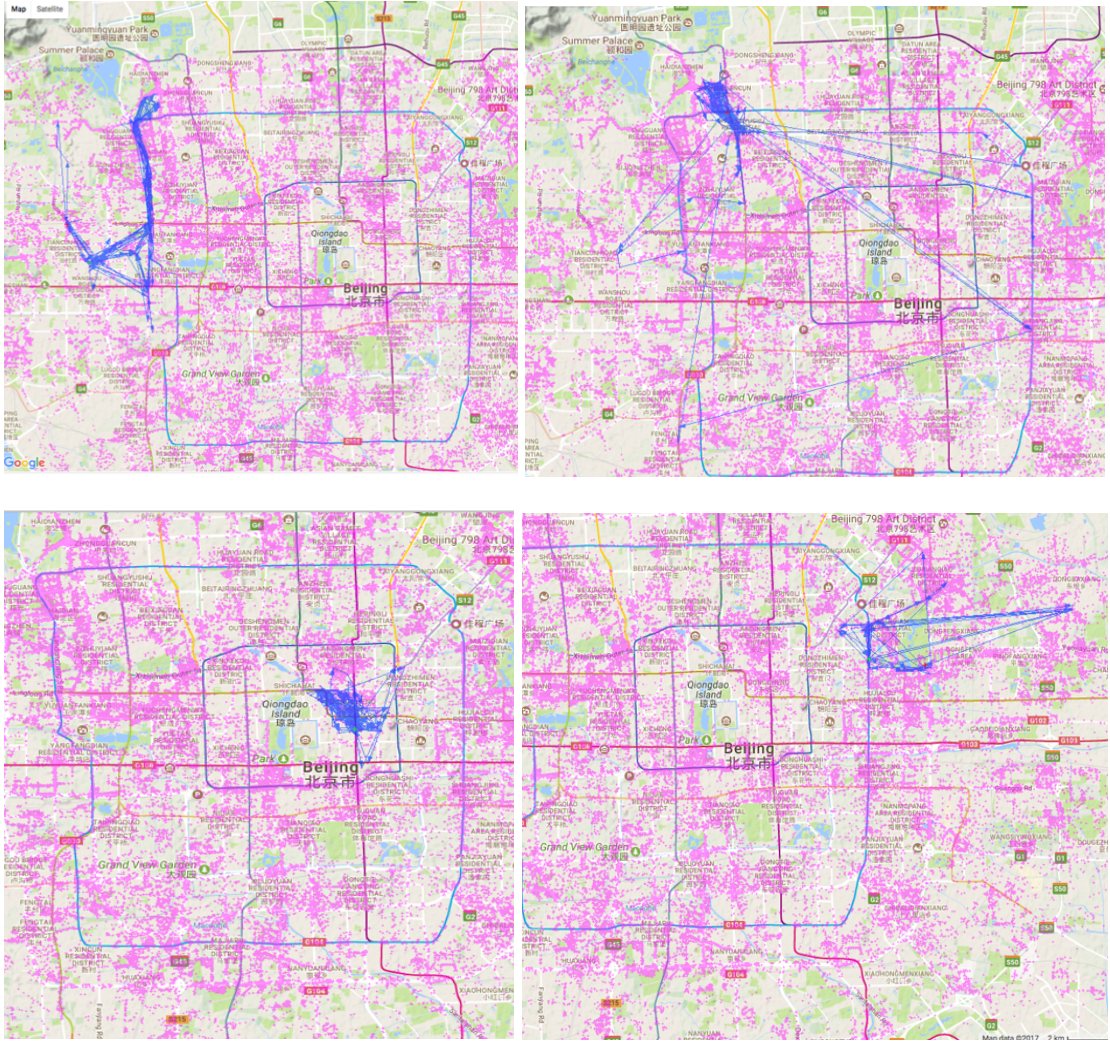

*Figure 3: Four bike sharing customers and their paths, overlaid on a map of Beijing. The pink shading portrays the relative density of contracts at locations across Beijing for the entire month of our dataset.*

Figure 3 visualizes four individual representative maps describing the path data for four high usage customers. We initially looked at variograms of paths and times of contracting for the most frequent customers of Mobike to see if there were specific periodic usages that could be elicited for them from the data. Typical paths for frequent customers are depicted in figure 3. Variograms are often used in tracking the movements of tagged animals for ecological and animal behavioral research. Such analyses were only possible for a few customers, because the number of rides during the one-month data period rapidly dropped below 500 rentals for individual riders; more extensive data may allow more powerful future analyses. Variograms are an unbiased way to visualize autocorrelation structure when migration, range shifting, drift, or other translations of the mean location are not happening. When drift occurs in the data, then the variogram represents a mixture of both the drift and the autocorrelation structure, each of which contains distinct movement behaviors.

Our exploratory statistics found the best fit for Mobike's top customer in this dataset was provided by then an Ornstein-Uhlenbeck model (Brownian motion restricted to a finite home range). Figure 4 shows the results of an analysis of Nyquist frequency, suggesting periodicities of 12 hours (2 x .25 x 24hours) and 7 days (2 x 3.5) for a two-harmonic analysis, and 8 hours for three-harmonic analysis.

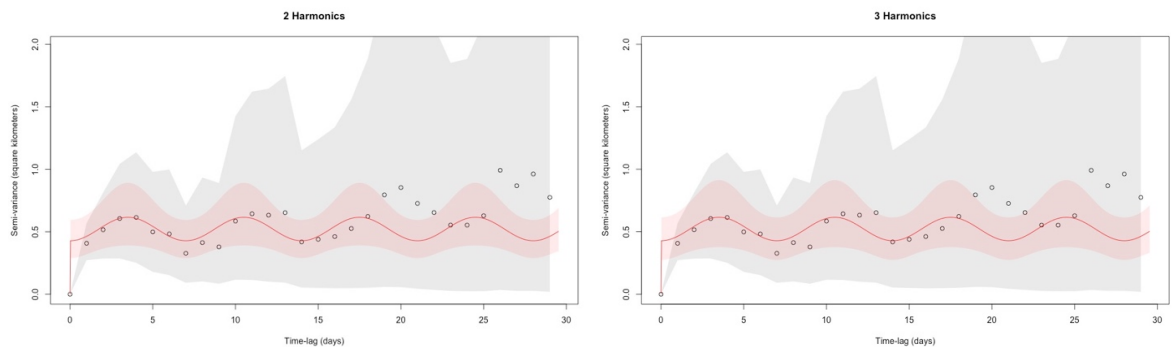

*Figure 4: Two and Three Period Nyquist Frequencies*

Similar regularities in behavior at around 8 hours, 12 hours and 7-day periods were seen in the top 10 customers. Such observations are hardly conclusive but highlight particular periodicities that we should probably be alert to (and around which we may want to adjust our models) for the entire dataset. The next section details how we binned time and location data into a $1024^3$ grid for the discrete wavelet transformation. The dataset grid divided into 1024 sections has approximately 40.8 minute time-steps where the Nyquist frequencies represent approximately 12, 18 and 247 step periodicities on that grid.

Figure 5 shows the Power Spectrum of a Morlet wavelet decomposition of the data flattened on time (i.e., total demand across Beijing for time-steps of 1 to 1024). Morlet wavelet decomposition provides a simple analysis for time-series data and is appropriate for eliciting periodicities. Note the black lines which highlight 'power ridges' at about 6 to 8 hours (8 to 12 steps, this black ridge wiggles up and down); about 12 hours (18 steps) and about 24 hours (36 steps). There appears to be higher power along a stripe at about the 7-day (247 steps) mark, but this is well into the shaded 'cone of influence' which excludes areas of edge effects. We do not have sufficient data to reveal 247 step periodicities (about ¼ of the entire time series), though our wavelets analyses using scales of 256 and 512 do suggest a power ridge at around that mark. The prominent power ridge at a periodicity of around 36 steps (36 * 40.8 minute steps is 24 hours) was not revealed in the Nyquist frequencies in the variograms of Mobike's most frequent customers.

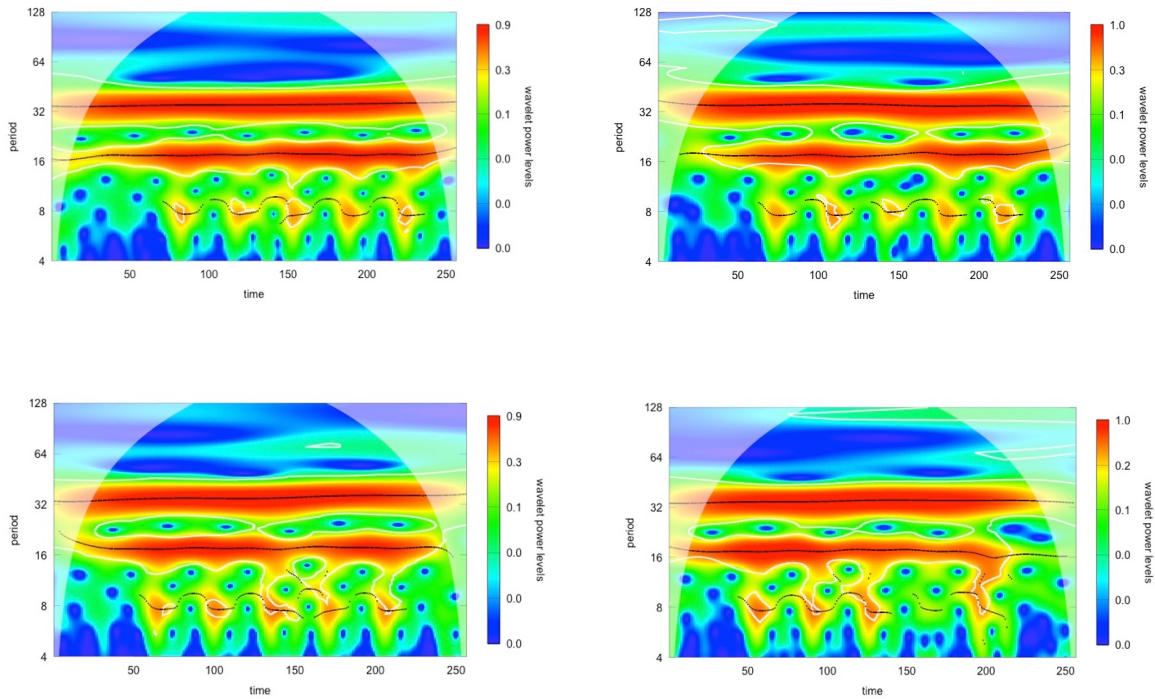

*Figure 5: Wavelet Power Spectrum for Flattened Dataset for each of the Four Time Slices: 1-256; 257-512;513-768;769-1024 steps (left-right, top-bottom). Each of the four represents about one-week and shows weekends and daily cycles prominently.*

Figure 5 clearly shows in week long intervals the 7, 12 and 24 hour periods of activity; these cycles are prominent even when data is restricted to particular sections of Beijing, and reflect the general daily cycles of Mobike's bike-sharing customers.

*Prediction of Bicycle Demand using Discrete Wavelet Transform with Thresholding*

Discrete wavelet transforms (DWT) expect to receive data in a regularized format, with a periodic recording of discrete locations. One can liken the space-time representation of [longitude, latitude, time] data to pixels forming a 3D image. DWT will analyze such data into scale and displacement coefficients for the mother wavelet and requires that the count of data on each dimension be the same power of 2 (the algorithm requires this, and we have chosen $2^{10} = 1024$). To remap the Mobike data in this fashion, data points were "binned" at a resolution of $\frac{1}{1024}$ on each dimension of the data; this comes to about [86.7 meters, 21.7 meters, 40.8 minutes] for the longitude, latitude and contract start time. We felt that this level of resolution represents an indifference point for most riders, as they would be willing to wait or walk to obtain a bike at less than this resolution in any dimension.

Thresholding acts as a low pass filter, filtering out noisy fluctuations in demand and carrying information about the major, stable features in a particular problem or situation. Mobike's demand appears to contain significant "noise" as the predictions without thresholding provide very little improvement over our baseline assumption that demand in epoch $i + 1$ will be exactly the same as demand in epoch $i$. But once these fine fluctuations in demand have been removed from the coefficients, we find stable demand behavior that can be successfully used in predicting future demand.

The validity of these assumptions can be assessed four our candidate periodicities, at 10, 18, 36 and 247 periods, revealed in the exploratory analysis about 6 to 8 hours (8 to 12 steps, this black ridge wiggles up and down so we chose 10); about 12 hours (18 steps); about 24 hours (36 steps); and at about 7 days (247 steps).

*Figure 6: Predictive Performance of Thresholded Level 3 to 8 Wavelet Decompositions*

| Period of 7 hours (10 steps) | | | | | | | | | |
|---|---|---|---|---|---|---|---|---|---|
| Level | Coeff | Compress | MSE | MAE | Skew | Kurt | MSE | MAE | Skew |
| 3 | 56 | 998,728 | 206.54 | 2.53 | 0.03 | 561.44 | 4.38% | -22.39% | -98.24% |
| 4 | 112 | 499,364 | 206.6 | 2.55 | 0.05 | 561.05 | 4.41% | -21.78% | -97.06% |
| 5 | 224 | 249,682 | 206.05 | 2.73 | 0.14 | 562.87 | 4.13% | -16.26% | -91.76% |
| 6 | 448 | 124,841 | 204.98 | 2.81 | 0.19 | 564.23 | 3.59% | -13.80% | -88.82% |
| 7 | 896 | 62,420 | 205.06 | 2.96 | 0.57 | 555.62 | 3.63% | -9.20% | -66.47% |
| 8 | 1792 | 31,210 | 197.87 | 3.26 | 1.7 | 556.83 | 0.00% | 0.00% | 0.00% |
| None | All | 0 | 197.87 | 3.26 | 1.7 | 556.83 | 0.00% | 0.00% | 0.00% |
| Period of 12 hours (18 steps) | | | | | | | | | |
| Level | Coeff | Compress | MSE | MAE | Skew | Kurt | MSE | MAE | Skew |
| 3 | 56 | 998,728 | 206.33 | 2.53 | 0.05 | 562 | 4.38% | -22.34% | -97.09% |
| 4 | 112 | 499,364 | 206.39 | 2.55 | 0.07 | 561.63 | 4.41% | -21.85% | -95.93% |
| 5 | 224 | 249,682 | 205.83 | 2.73 | 0.16 | 563.48 | 4.12% | -16.22% | -90.70% |
| 6 | 448 | 124,841 | 204.78 | 2.8 | 0.21 | 564.78 | 3.59% | -14.03% | -87.79% |
| 7 | 896 | 62,420 | 204.85 | 2.96 | 0.59 | 556.23 | 3.63% | -9.28% | -65.70% |
| 8 | 1792 | 31,210 | 197.68 | 3.26 | 1.72 | 557.43 | 0.00% | 0.00% | 0.00% |
| None | All | 0 | 197.68 | 3.26 | 1.72 | 557.43 | 0.00% | 0.00% | 0.00% |
| Period of 24 hours (36 steps) | | | | | | | | | |
| Level | Coeff | Compress | MSE | MAE | Skew | Kurt | MSE | MAE | Skew |
| 3 | 56 | 998,728 | 186.06 | 2.25 | 0.38 | 659.61 | 2.81% | -24.42% | -83.11% |
| 4 | 112 | 499,364 | 186.08 | 2.27 | 0.4 | 659.34 | 2.82% | -23.95% | -82.22% |
| 5 | 224 | 249,682 | 186.87 | 2.46 | 0.54 | 653.49 | 3.25% | -17.34% | -76.00% |
| 6 | 448 | 124,841 | 187.16 | 2.55 | 0.65 | 648.86 | 3.42% | -14.54% | -71.11% |
| 7 | 896 | 62,420 | 186.61 | 2.69 | 0.99 | 640.35 | 3.11% | -9.91% | -56.00% |
| 8 | 1792 | 31,210 | 180.98 | 2.98 | 2.25 | 638.06 | 0.00% | 0.00% | 0.00% |
| None | All | 0 | 180.98 | 2.98 | 2.25 | 638.06 | 0.00% | 0.00% | 0.00% |
| Period of 7 days (247 steps) | | | | | | | | | |
| Level | Coeff | Compress | MSE | MAE | Skew | Kurt | MSE | MAE | Skew |
| 3 | 56 | 998,728 | 167.29 | 2.01 | 3.44 | 721.14 | 2.19% | -22.02% | -33.72% |
| 4 | 112 | 499,364 | 167.34 | 2.02 | 3.46 | 720.72 | 2.23% | -21.59% | -33.33% |
| 5 | 224 | 249,682 | 167.97 | 2.18 | 3.57 | 715.12 | 2.61% | -15.69% | -31.21% |
| 6 | 448 | 124,841 | 168.21 | 2.24 | 3.66 | 710.5 | 2.76% | -13.20% | -29.48% |
| 7 | 896 | 62,420 | 167.93 | 2.35 | 3.96 | 701.68 | 2.58% | -8.99% | -23.70% |
| 8 | 1792 | 31,210 | 163.7 | 2.58 | 5.19 | 698.92 | 0.00% | 0.00% | 0.00% |
| None | All | 0 | 163.7 | 2.58 | 5.19 | 698.92 | 0.00% | 0.00% | 0.00% |

Notice that in figure 6 the MSE only slightly, but steadily, changes with higher thresholds, where each step up in threshold level doubles the number of coefficients containing information about the bicycle demand. But the other statistics describe a more nuanced story. Discrete wavelet transforms keep the average error close to zero, but the mean absolute error deteriorates as threshold level goes up. The reason being that errors are very keratotic and tend to be right-skewed (the right-hand, positive tail of the distribution is extended). Skewness and mean absolute error decline with lower threshold levels. We can conclude from this that the impact of 'noise' on our estimates is to introduce bias and dispersion to the predictions, even though accuracy of mean estimation (in a squared-error loss context) is slightly improved (2-4%) with a move to higher thresholding, but the corresponding improvements in mean absolute error show 20-25% improvement; and skewness statistics show 30-35% improvement. Thresholding down to level 3 compresses the explicit representation of the information in the dataset by a factor of nearly 1 million, with only a 2-4% information loss (for predictive purposes) in our focal MSE performance measure.

The full DWT "overfitted" the data, reflecting the fact that this data contains numerous infrequent users whose behavior is not well predicted by our model (as illustrated in figure 2 Gini and Lorenz plots). Within the context of the DWT analysis, this is 'noise' rather than 'signal', and it is useful look at skewness and kurtosis to better understand the source of errors outside our DWT. What we see is a reduction in skewness; 'noise' right-skews the predictions (the predictions require that we deliver more bikes than are needed to a specific location) and increase the over and under-delivery of bikes at specific locations. Thresholding filters out this right-skewing spurious customer behavior caused by the many infrequent users distorting demand and focuses managerial policy and procedures on the more frequent and more predictable customers.

*Analysis of Predictions after Threshold "Signal" Filtering*

In this section we graph our demand predictions on a period by period basis, for various lags, to visually inspect the stability of our results. This also allows inspection of DWT predictions that are not restricted to lead times of 10,18,36 and 247 periods as in the prior section. The move to a comprehensive analysis of cycle times creates a relatively dense output, so we will use graphs to visualize that signal biases that are removed or retained in our analysis.

Behavioral cycles conjectured at the start of this study were found to be predictable and uniform across the dataset. It is easier to visualize smaller subsamples of our more than 55 million datapoints, and these are presented below; though the reader should recognize that we have reviewed the entire dataset where these results more generally are found to be representative. Figure 21 summarizes the average error in demand prediction across the 2-dimensional 1024 x 1024 location grid when the prediction from the inverse wavelet transform is applied to a time period that leads the predicted time by a number from 1 to 500. Here we are looking for human behavioral cycles that are equal to that lead time (1 to 500) shown on the lead axis. We repetitively apply that predict demand at time periods 1 to 512 (we chose half the full-time length of the data, because a graph from 1 to 1024 looks more or less like an uncut lawn and is difficult to interpret). We have retained the raw errors in these graphs (rather than MSE) because there is information in the fact that some errors are negative (they overestimate demand) while others are positive. We can vaguely identify the 10, 18 and 36 period cycles towards the zero axis, and multiples of these regularly appear as we increase to larger and larger lead times. Most pronounced is an apparent phase-shifting where these cycles seem to travel across diagonals (note that the lead times start counting at the period axis index, so this is not an artifact of improper graphing). This phase shifting appears to be conserved across all threshold levels (figure 22 shows the graph for predictions from the inverse of the threshold level 3 of the DWT).

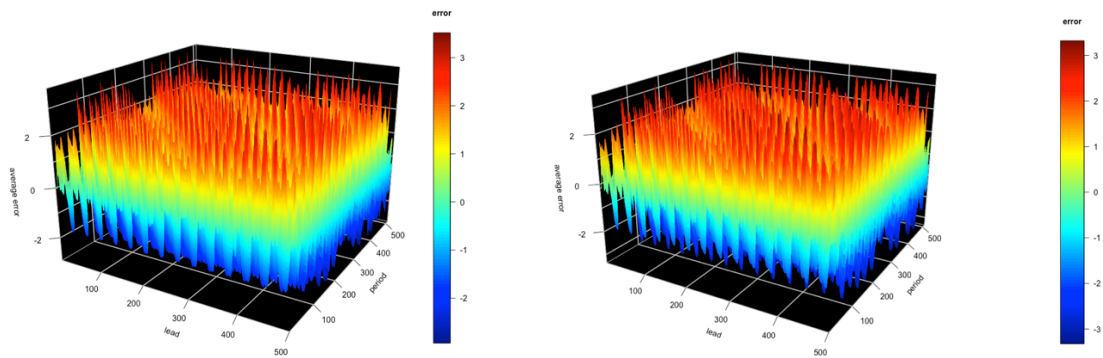

*Figure 7: Average Error of Wavelet **Threshold 8** (left) and **Threshold 3** (right) Prediction for 1 to 500 Periods Lead into the Future, for the first 512 Periods*

We can focus our analysis is smaller lead times order to determine the fine effects of thresholding. Comparing figure 23 (threshold level 8) to figure 24 (threshold level 3) we see that thresholding substantially smooths the predictions in to evenly spaced rounded mountains, both on the lead and period axes. Our detailed investigation of specific periods showed that this smoothing removed to different sources of 'noise': (1) short periods of reduction in bike demand due to poor air quality or bad weather; and (2) short periods of abnormal demand from infrequent riders. One might argue that air quality and weather could be forecast up to about 7 days, and in future models could augment predictions through a very simple additive model. One would compute average reductions in overall demand for a particular type of weather, and then adjust our prediction with a fixed factor computer from this computation.

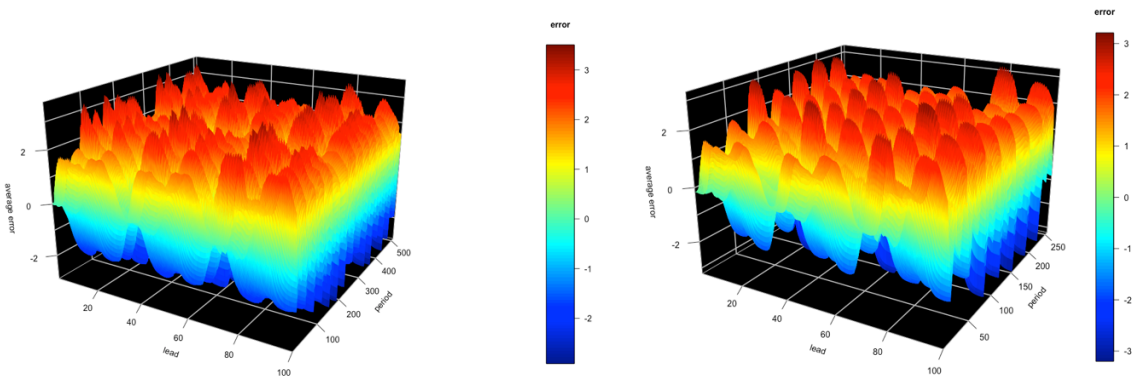

*Figure 8: Average Error of Data Reconstructed from Wavelet **Threshold Level 8** (left) and **Threshold Level 3** (right) Prediction for 1 to 100 Periods Lead into the Future, for the first 512 Periods*

Our thresholding adjusts the prediction models for three distinct types of noise that arise from overfitting of the data:

1) Dispersion noise for demand at a location
2) Right-skewed bias noise for demand at a location
3) Phase-shift bias noise for time of demand

Figure 8 tells much of the story about the nature of 'signal' extraction performed by thresholding. Notice that the peaks and valleys are much smoother at level 3 than at level 8, and also level 3 error values sit directly behind one another. The diagonal valleys on the time (period) axis on figure 7 suggest that DWT not only adds location noise in the form of right-skewed predictions and higher variance; but that this noise is also added on the time axis in what appears as phase-shifting. Since DWT and thresholding are symmetric on all three axes, phase-shifting is an expected type of noise that would appear in overfitted models. Lowering the threshold filters out phase-shift noise. It also filters out inconsistencies in prediction that result from overfitting; these arise from dispersion-based noise that shows up as jagged peaks and valleys (looking along the 'lead' axes in figure 8) in predictions for a particular period; as well as inconsistent predictions for different periods (looking along the 'period' axes in figure 8) when the dataset is recovered from higher level thresholding of the DWT. All of these reflect specific types of prediction 'noise' arising from infrequent users of Mobike's bike-sharing service.

## Discussion

Consumers are creatures of habit, and we relied on this trait in pursuit of our research objective: to identify human behavioral habits reflected in demand cycles which can be used for the improved geographical positioning of Mobike bikes in a predefined service region. Demand prediction is a core challenge to companies such as Mobike and motivates programs and managerial decision making.

We used exploratory statistics from ecological studies called variograms to identify the best fit for Mobike's top customers, finding a good fit for an Ornstein-Uhlenbeck model with periodicities of 12 hours (2 x .25 x 24hours) and 7 days (2 x 3.5) for a two-harmonic analysis, and 8 hours for three-harmonic analysis. Similar regularities in behavior at around 8 hours, 12 hours and 7-day periods were seen in the top 10 customers, and we used these cycles as a basis for evaluating more computationally intensive wavelet decompositions.

The central model developed in this research used a discrete wavelet transforms (DWT) to analyze the space-time tensor of [longitude, latitude, time] data into scale and displacement coefficients for a Daubechies mother wavelet. We "binned" the Mobike data at an 'indifference' resolution of [86.7 meters, 21.7 meters, 40.8 minutes] for the longitude, latitude and contract start time.

We were able to represent most of the relevant information in the dataset in a set of threshold level 3 coefficients that was around $\frac{1}{1,000,000}$th the size of the original data (56 DWT coefficients vs. ~55 million datapoints). This indicates that bike-sharing demand behavior is exceptionally strong and regular and can be elicited through thresholded DWTs. Even at the maximum level of compression, we were easily able to validate the accuracy of the main human behavioral bike-sharing cycles of 7 hours (10 steps); 12 hours (18 steps); 24 hours (36 steps); and 7-day (247 steps) that we postulated from variograms. We surmised that 'noise' from: (1) short periods of reduction in bike demand due to poor air quality or bad weather; and (2) short periods of abnormal demand from infrequent riders were the main cause of errors introduced into our predictions of demand from the induced human cyclical demand behavior. Both could potentially be ameliorated by separate models that could then be added to the results of thresholded DWT demand predictions.

The resulting models of behavior would substantially improve over predictions based on the raw data, which were shown to be substantially the same as those for DWT thresholded at level 8. The full DWT "overfitted" the data, reflecting the fact that this data contains numerous infrequent users whose behavior may be unstable and poorly predicted by models that assume habitual behavior (as illustrated in figure 2 Gini and Lorenz curves). Thresholding filters out spurious customer behavior and anchors demand cycles and prediction on the behavior of the more frequent, profitable and predictable customers. Managerial strategy could potentially try to achieve the same effect (and greater predictability of the location of demand for bikes) by tracking the activities, as well as incentivizing high-frequency users. Such programs are common in consumer industries such as cinema, restaurants and so forth.

The current study elicits an accurate model for the habitual cyclical demand embedded in human behavior, using only the minimal data that is available from the inexpensive geo-tracking hardware built into Mobike's bike sharing systems. We were limited in this research to one month of data, which was probably too short for fully exploiting wavelets methods for prediction given the periodicities that we saw in the demand data. Future studies will extend the time-span of data, and expand the model to account for weather, air quality and other factors that have been shown in prior studies to influence demand.